\documentclass[fleqn,usenatbib]{mnras}

\usepackage{newtxtext,newtxmath}

\usepackage[T1]{fontenc}

\DeclareRobustCommand{\VAN}[3]{#2}
\let\VANthebibliography\thebibliography
\def\thebibliography{\DeclareRobustCommand{\VAN}[3]{##3}\VANthebibliography}

\usepackage{graphicx}	
\usepackage{amsmath}	
\usepackage{pdflscape}
\usepackage{siunitx}

\title[Laboratory 4f and 5d Fe II energy levels]{Laboratory confirmation and improved Accuracy of 4f and 5d energy levels of Fe II previously identified from stellar spectra}

\author[M. Ding et al.]{
Milan Ding,$^{1}$\thanks{E-mail: milan.ding15@imperial.ac.uk}
Hiroki Kozuki,$^{1}$
Florence Concepcion,$^1$
Gillian Nave$^{2}$
and Juliet C. Pickering$^{1}$
\\
$^{1}$Blackett Laboratory, Imperial College London, London SW7 2AZ, UK \\
$^{2}$National Institute of Standards and Technology, Gaithersburg, MD 20889-8422, USA
}

\date{Accepted XXX. Received YYY; in original form ZZZ}

\pubyear{2024}

\begin{document}
\label{firstpage}
\pagerange{\pageref{firstpage}--\pageref{lastpage}}
\maketitle

\begin{abstract}
    Many energy levels of singly ionised iron (Fe~II, $Z=26$) remain uncertain or experimentally unknown. Their identification and spectral line data are required in reliable astrophysical spectral analyses. In motivation for improving the atomic data of Fe~II, we analysed emission spectra of a Fe-Ne plasma produced by a Penning discharge lamp recorded by high-resolution Fourier transform spectroscopy in the region 9000 -- 27,000~cm$^{-1}$ (11,111 -- 3704~\AA{}). Semi-empirical transition probability calculations and stellar spectra of Fe~II were used to guide the analysis. In total, 24 energy levels of the 3d$^6$4f and 3d$^6$5d configurations of Fe~II lying between 122,351 -- 127,881~cm$^{-1}$ were confirmed in the laboratory for the first time, in agreement with their identities proposed by previous investigations involving only stellar spectra. Level energy and wavelength uncertainties of the 24 levels are improved by up to an order-of-magnitude compared to previously published values. These results will enable more reliable application of Fe~II in astrophysical spectroscopic analyses and support further investigations of the spectrum and energy levels of Fe~II.
\end{abstract}




\section{Introduction}
The high relative cosmic abundance and complex spectra of the iron-group elements contribute to most of the opacities and spectral lines in many astrophysical objects, such as stars, nebulae, and galaxies. Singly ionised iron (Fe~II, $Z=26$) is one of the main species of interest. Reference spectral line data for astrophysical spectral synthesis are never complete and always subject to improvement. For Fe~II,  the current challenge arises from closely packed highly excited energy levels near the ionisation limit (e.g. levels of the 3d$^6$4f and 3d$^6$5d configurations, referred to as the 4f and 5d levels), requiring observation of the Fe II spectrum under a variety of plasma conditions and extensive parameterisation of semi-empirical atomic structure calculations \citep[e.g.][]{kurucz2017including, cowan1981theory}.

\cite{castelli2010new} had identified 109 4f and 14 (4, 5, and 6)d energy levels of Fe II using only high-resolution spectra of the Fe-overabundant chemically peculiar star HR~6000 recorded by the UVES instrument at the VLT. Soon after, the most extensive and recent laboratory investigation of the spectrum and energy levels of Fe~II by \cite{nave2012spectrum} confirmed roughly half of the levels identified by \cite{castelli2010new}.

In this paper, we present identification of 24 energy levels of the 4f and 5d singly excited configurations of Fe~II that were proposed by analysing stellar spectra \citep{castelli2010new} but were unable to be confirmed by laboratory spectroscopy of Fe~II by \cite{nave2012spectrum}. 
In total, 125 transitions within 5072 -- 7868~\AA{} were observed and classified in the identification of these 24 levels, with level energies typically accurate to 0.01~cm$^{-1}$. New Ritz wavelengths are typically accurate to 0.0027~\AA{}.



\section{Experimental Details}
\subsection{Fourier transform spectroscopy}
The spectra used in the present work had been recorded by Fourier transform (FT) spectroscopy for the ongoing spectrum and energy level analysis of Fe~III at Imperial College London \citep{pickering2019recent,concepcion2022spectrum}. A Penning discharge lamp \citep[PDL, e.g.][]{heise1994radiometric} with neon carrier gas and pure Fe cathodes (99.9\%) was used with the aim to populate Fe~III energy levels in an Fe-Ne plasma. Consequently, many high-lying levels of Fe~II were also populated and hence their transitions were observed. For this reason, these PDL spectra are advantageous for studying the high-lying 4f and 5d Fe~II levels compared to the lower effective temperature Fe hollow cathode lamp used for FT spectroscopy of Fe~II by \cite{nave2012spectrum}. 

Two spectra of the Fe-Ne PDL were recorded using the 2-m National Institute of Standards and Technology (NIST) IR-UV FT spectrometer: spectrum FE1120.2 covers the wavelength range 5000 -- 11,111~\AA{} (20,000 -- 9000~cm$^{-1}$) and spectrum FE1118.6 covers a shorter wavelength range 3704 -- 6250~\AA{} (27,000 -- 16,000~cm$^{-1}$), and their experimental parameters are summarised in Table~\ref{tab1}.
\begin{table*}
	\centering
	\caption{Experimental parameters of the two Fe-Ne PDL FT Spectra.}
	\label{tab1}
    \setlength{\tabcolsep}{5pt}
	\begin{tabular}{ccccccccccc} 
    \hline
    Spec. No. & Date & Spec. Range & Resolution & PDL Current & PDL Pressure & Detector & Optical Filter & Co-adds & $k_{\text{eff}}$ \\
           &      & (cm$^{-1}$) & (cm$^{-1}$) & (A) & (mbar) & & & & ($10^{-7}$) \\
    \hline
    FE1120.2 & 20 Nov. 1998 & 9000 -- 20,000 & 0.04 & 1.70 & $5.5\times10^{-4}$ & Si diode & OG530 & 50 & \multicolumn{1}{r}{$0.748\pm0.140$} \\
    FE1118.6 & 18 Nov. 1998 & 16,000 -- 27,000 & 0.08 & 1.75 & $4.2\times10^{-4}$ & Si diode & CuSO$_4$ & 50 & $-1.438\pm0.232$ \\
    \hline
	\end{tabular}
\end{table*}
The spectral resolution of each spectrum was chosen as the minimum resolving power required to observe no instrumental ringing on the lowest wavenumber spectral line, so that the resolution is Doppler limited. Wavenumber, signal-to-noise ratio (S/N), relative intensity (area-under-line-profile), and full-width-at-half-maximum (FWHM) were obtained by fitting a Voigt profile to each observed spectral line using the program XGREMLIN \citep{nave2015xgremlin} and tabulated as a line list. In total, 2943 and 2312 lines were fitted in the spectra FE1120.2 and FE1118.6, respectively. The statistical wavenumber uncertainty was estimated using fitted parameters as
\citep{davis2001fourier}
\begin{equation}\label{eq:stat_unc}
    \Delta\sigma_{\text{obs}}\approx\frac{1}{\sqrt{N}}\frac{\text{FWHM}}{\text{S/N}},
\end{equation}
where $N$ is the number of points within the FWHM of a line and S/N is capped at 100 to prevent unrealistic uncertainties as small deviations (i.e. fit residuals) from the ideal Voigt profile generally exist around 1\% of the S/N of an observed spectral line.

\subsection{Wavenumber and relative intensity calibration}
The two Fe-Ne PDL FT spectra were wavenumber calibrated using the wavenumber calibration factors $k_{\text{eff}}$ in Table~\ref{tab1}, defined as \citep{davis2001fourier,nave2004reference}
\begin{equation}\label{eq:wn_cal}
    \sigma\textsubscript{cal} = (1+k\textsubscript{eff})\sigma\textsubscript{obs},
\end{equation}
where $\sigma\textsubscript{obs}$ is the observed wavenumber from spectral line fitting and $\sigma\textsubscript{cal}$ is the wavenumber calibrated onto the absolute scale. Both spectra were wavenumber calibrated by their respective $k_{\text{eff}}$ values determined using Fe~II Ritz wavenumbers and Ritz wavenumber uncertainties of \cite{nave2012spectrum}, accessed also through the NIST Atomic Spectra Database \citep{kramida2023nist}. 

In total, 94 and 45 Fe~II lines were used to calibrate FE1120.2 and FE1118.6, respectively. These lines have S/Ns above 20 with no features of line blending and self-absorption. The large number of calibration lines can lead to underestimated calibration uncertainties in the two Fe-Ne PDL FT spectra. To avoid this, and to ensure correlation between calibration uncertainties of the two Fe-Ne PDL FT spectra and those of the spectra in \cite{nave2012spectrum}, the global calibration factor uncertainty of $4\times10^{-8}$ from \cite{nave2012spectrum} was added to $\Delta k_{\text{eff}}$. Therefore, $(\Delta k_{\text{eff}} + 4\times10^{-8})\sigma_{\text{obs}}$ was the wavenumber calibration uncertainty added in quadrature with the statistical wavenumber uncertainty to yield the final observed wavenumber uncertainty of each line.

The two Fe-Ne PDL FT spectra were also intensity calibrated onto the same relative scale corresponding to relative photon flux using the program XGREMLIN, where instrumental response functions were obtained from tungsten lamp continuum spectra. Uncertainties in spectral line relative intensities are expected to be at least 20\% for the Fe~II lines newly classified in this work due to low S/Ns (about 10 on average). Uncertainties in relative intensity, for blended lines, and when comparing lines observed in FE1118.6 and FE1120.2, are also more uncertain, the latter is due to different instrumental conditions (e.g. see currents and presssures of Table~\ref{tab1}). Therefore, the relative intensity values should be considered only as guides.

\section{Analysis and results}
\subsection{Energy levels and classified transitions}
All 4f and 5d Fe~II energy levels identified by \cite{castelli2010new} using stellar spectra of HR~6000, but unconfirmed by \cite{nave2012spectrum}, were investigated. The branching fractions of the transitions from each of these levels calculated by R. L. Kurucz in his most recent gf2601.pos file \citep[][]{kurucz2013fe} from his database \citep{kurucz2017including} were compared with experimental relative line intensities. An energy level was considered potentially identified if two or more of its spectral lines calculated with the highest transition probabilities were observed in the Fe-Ne PDL FT spectra at the expected relative intensities. Then, if the observed wavenumbers of these lines agree with their optimised Ritz values within wavenumber uncertainties, the level was concluded to be confirmed. 

Level energies and Ritz wavenumbers were optimised using the program LOPT \citep{kramida2011program}. Observed wavenumbers and wavenumber uncertainties of transitions of the confirmed levels were grouped by FE1120.2 and FE1118.6 in the input for LOPT to include calibration uncertainties in the estimation of uncertainties in optimised level energies and Ritz wavenumbers. Wavenumbers obtained from spectral lines with unreliable Voigt profile fitting, e.g., noise induced anomalous FWHM or significant blending, were weighted zero in the level optimisation. Energies of the lower levels connecting to the optimised 4f and 5d levels were fixed using their previously published values and uncertainties \citep{nave2012spectrum,kramida2023nist}. The list of 4f and 5d Fe~II energy levels successfully confirmed in the laboratory for the first time and their observed spectral lines are presented in Table~\ref{tab2} and Table~\ref{tab3}, respectively. 
\begin{table*}
	\centering
	\caption{Observed energy levels of the 3d$^6$4f and 3d$^6$5d configurations of Fe~II.}
	\label{tab2}
	\begin{tabular}{cccccccc} 
    \hline
    Sub-configuration$^a$ & Term$^a$ & $J$$^a$ & $E$$^b$ & $\Delta E$$^b$ & $N_{\text{lines}}$$^c$ & $E_{\text{CK}}$$^d$ & \multicolumn{1}{c}{$E$ - $E_{\text{CK}}$}\\
     & & & (cm$^{-1}$) & (cm$^{-1}$) & & \multicolumn{1}{c}{(cm$^{-1}$)} & \multicolumn{1}{c}{(cm$^{-1}$)}   \\
    \hline
    3d$^6$($^3$P2)4f & 2[3]$^\circ$ &7/2 &122351.535   &0.025  &4(2)&122351.488  &0.047 \\  
    3d$^6$($^3$P2)4f & 2[5]$^\circ$ &11/2&122351.849  &0.008   &2(1)&122351.810 & 0.039 \\
    3d$^6$($^3$H)4f  & 6[4]$^\circ$ & 7/2  &122980.411	&0.018 &5(3)&122980.408 & 0.003\\
    3d$^6$($^3$H)4f  &5[5]$^\circ$ & 9/2  &123269.391	&0.007 &7(3)&123269.378 & 0.013\\
    3d$^6$($^3$H)4f  &4[3]$^\circ$ & 7/2  &123451.511	&0.011 &6(3)&123451.449 & 0.062\\
    3d$^6$($^3$F2)4f &4[4]$^\circ$ & 7/2 &124385.029	&0.016 &4(2)&124385.010 & 0.019\\
    3d$^6$($^3$F2)4f &4[5]$^\circ$ & 9/2 &124385.745	&0.021 &6(4)&124385.706 & 0.039\\
    3d$^6$($^3$F2)4f &4[4]$^\circ$ & 9/2 &124401.963	&0.014 &6(5)&124401.939 & 0.024\\
    3d$^6$($^3$F2)4f &4[6]$^\circ$ & 11/2&124402.559	&0.011 &9(3)&124402.557 & 0.002\\
    3d$^6$($^3$F2)4f &3[5]$^\circ$ & 11/2&124626.958	&0.008 &8(5)&124626.900 & 0.058\\
    3d$^6$($^3$F2)4f &3[5]$^\circ$ & 9/2 &124636.141	&0.026 &5(3)&124636.116 & 0.025\\
    3d$^6$($^3$F2)4f &3[3]$^\circ$ & 7/2 &124642.017	&0.010 &5(3)&124641.989 & 0.028\\
    3d$^6$($^3$F2)4f &3[6]$^\circ$ & 11/2&124656.554	&0.011 &4(2)&124656.535 & 0.019\\
    3d$^6$($^3$F2)4f &2[4]$^\circ$ & 7/2 &124783.783	&0.019 &4(3)&124783.748 & 0.035\\
    3d$^6$($^3$F2)4f &2[4]$^\circ$ & 9/2 &124793.947	&0.011 &6(3)&124793.905 & 0.042\\
    3d$^6$($^3$F2)4f &2[5]$^\circ$ & 11/2&124803.975	&0.005 &4(1)&124803.873 & 0.102\\
    3d$^6$($^3$F2)4f &2[5]$^\circ$ & 9/2 &124809.758	&0.007 &3(2)&124809.727 & 0.031\\
    3d$^6$($^3$G)4f  &5[6]$^\circ$ & 13/2 &127489.513	&0.007 &3(0)&127489.429 & 0.084\\
    3d$^6$($^3$G)4f  &4[4]$^\circ$ & 9/2  &127869.923	&0.008 &5(1)&127869.892 & 0.031\\
    3d$^6$($^3$G)4f  &4[6]$^\circ$ & 11/2 &127880.453	&0.008 &7(5)&127880.436 & 0.017\\
    3d$^6$($^3$H)5d  &$^4$I & 15/2        &124357.282	&0.003 &6(3)&124357.304 & -0.022\\
    3d$^6$($^3$H)5d  &$^4$K & 13/2        &124415.367	&0.005 &8(4)&124415.353 & 0.014\\
    3d$^6$($^3$H)5d  &$^2$I  & 11/2       &124976.009	&0.005 &4(2)&124976.008 & 0.001\\
    3d$^6$($^3$F2)5d &$^4$H  & 13/2      &125732.993	&0.008 &4(3)&125732.991 & 0.002\\
    \hline
    \multicolumn{8}{l}{$^a$ Designations following \cite{kurucz2013fe}} \\
    \multicolumn{8}{l}{$^b$ Optimised level energy and its standard uncertainty} \\
    \multicolumn{8}{l}{$^c$ No. of lines observed (no. of weak or blended lines weighted zero in the optimisation)} \\
    \multicolumn{8}{l}{$^d$ Level energies from \cite{castelli2010new}}\\
	\end{tabular}
\end{table*}
Term designations of the 4f and 5d levels were made using $JK$-coupling and $LS$-coupling labels from \cite{kurucz2013fe}, respectively. The spectral lines of Table~\ref{tab3} are in ascending order of upper level energy and, for each upper level, in descending order of weighted transition probability $gA$ obtained from the $\log(gf)$s of \cite{kurucz2013fe}.

\subsection{Analysis of weak and blended lines}
As previously mentioned, the average S/N of the transitions observed is about 10. In general, such low S/N lines constitute the majority of entries in a line list, and frequent blending of weak lines pose great challenges for empirical analyses of complex atomic spectra. This was no different for the present work in confirming levels that had originally been identified by \cite{castelli2010new} who used stellar spectra of HR~6000. For weak and blended lines in our laboratory spectra, spectral line fitting becomes ambiguous and lines could be incorrectly judged to be noise or non-blended. Such lines often cannot be fitted reliably using the Voigt profile and their wavenumbers are weighted zero in the energy level optimisation. 

For each energy level proposed by \cite{castelli2010new}, every single one of its spectral lines that was expected to have sufficient transition probability \citep{kurucz2013fe} to be observable above the noise level of the Fe-Ne PDL FT spectra were searched for in these laboratory spectra and analysed. For example, lines concluded to be reliably identified had relative intensities agreeing with calculated transition probabilities after factoring in blending, had wavenumbers agreeing within uncertainties with Ritz wavenumbers, and did not correspond to any experimentally known spectral line of Fe~I-III and Ne~I-III \citep{kramida2023nist}. Very weak spectral lines that could not be fitted reliably, but were used in identifying the energy levels, are included in Table~\ref{tab3}. The comment column of Table~\ref{tab3} summarises the situation for each line.

\begin{figure*}
	\includegraphics[width=\linewidth]{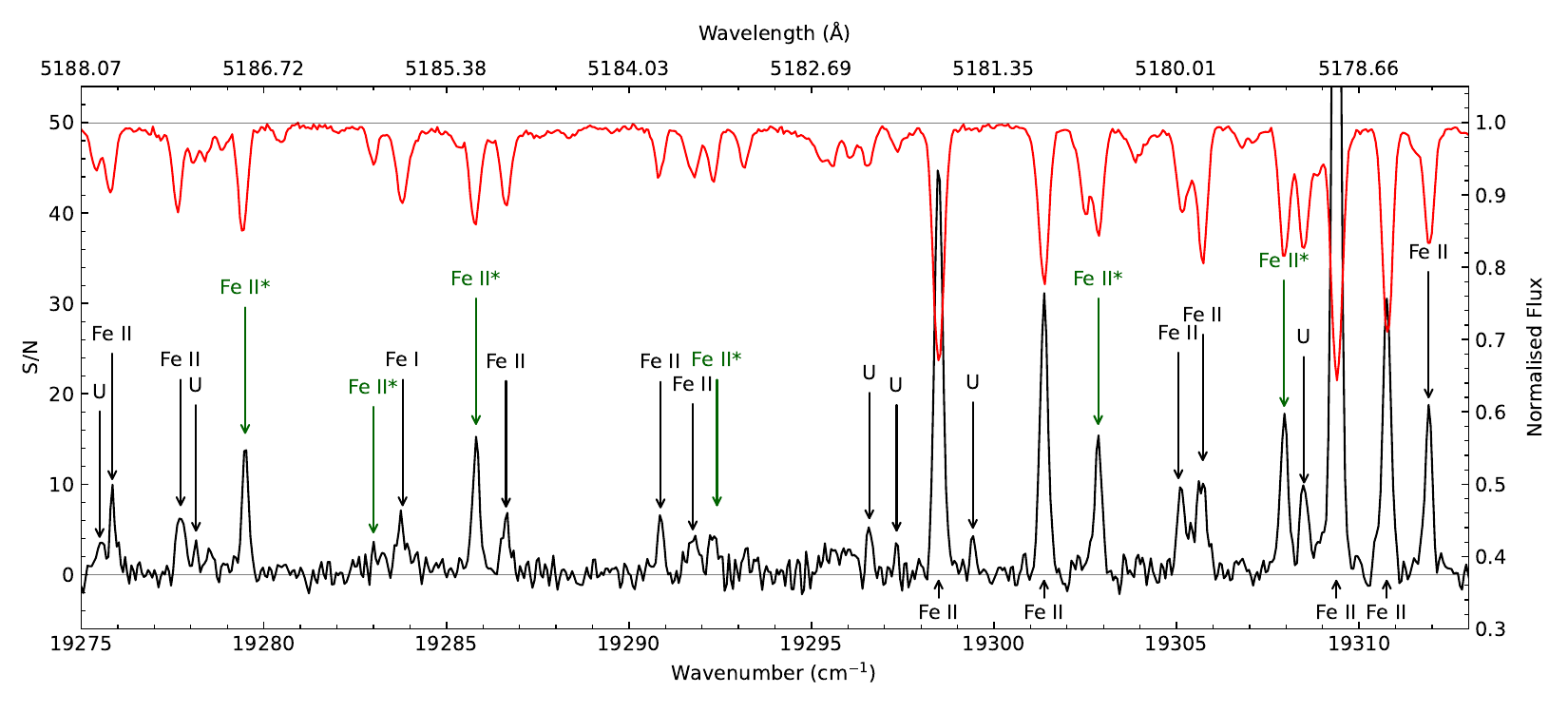}
    \caption{Section of Fe-Ne PDL FT emission spectrum FE1118.6 in the S/N scale (black) and the corresponding section of the UVES high-resolution stellar absorption spectrum of star HR~6000 in the normalised flux scale (red). Wavenumbers and wavelengths are vacuum values. Lines with arrowed labels Fe~I, Fe~II, and U (unclassified) were deduced using experimentally classified line lists \citep{nave1994new,nave2012spectrum,kramida2023nist}. All labelled spectral lines were fitted with Voigt profiles for the line list, those labelled as Fe~II* are newly experimentally confirmed in the present work.}
    \label{fig1}
\end{figure*}
\subsection{Accuracy improvements}
Figure~\ref{fig1} shows a comparison between one section of the high-resolution spectra of the star HR~6000 used by \cite{castelli2010new} and the high-resolution Fe-Ne PDL FT spectrum FE1118.6. The spectrum of the star HR~6000 in Figure~\ref{fig1} was used for the analyses of \cite{castelli2010new, castelli2009new, castelli2007refined}, recorded on the 19th of September 2005 under program 076.D-0169 and available at the ESO science archive \citep{romaniello2024eso}. At 19,000~cm$^{-1}$ (5260~\AA{}), the resolving power of the Fe-Ne PDL FT spectrum FE1118.6 (237,500) is more than twice the resolving power used by UVES (107,200) to obtain the stellar spectrum. Therefore, one could expect at least a factor of two improvement in level energy and wavelength accuracy. However, this is not certain as the S/N and line widths are likely higher and larger in the stellar spectrum from the comparison shown in Figure~\ref{fig1}.

\begin{figure}
	\includegraphics[width=\linewidth]{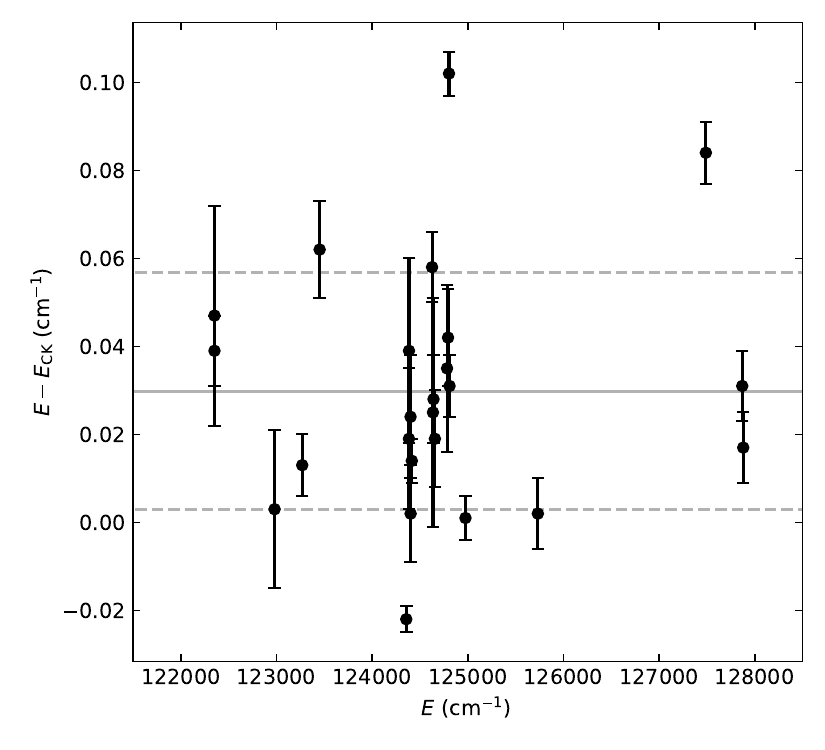}
    \caption{Difference between optimised level energies determined in the present work and those determined in \citet{castelli2010new}. The error bars show uncertainties of the optimised level energies of the present work, the solid horizontal line shows mean deviation, and the dashed horizontal lines show plus or minus the root-mean-square deviation from the mean deviation.}
    \label{fig2}
\end{figure}

In the present work, uncertainties of optimised level energies are on average 0.012~cm$^{-1}$ and Ritz wavenumbers are accurate up to 0.003~cm$^{-1}$. The uncertainties of the level energies and wavelength were not estimated in \cite{castelli2010new}. However, a comparison is possible using deviations between \cite{castelli2010new} level energies and corresponding optimised values of the present work, and this is shown in Figure~\ref{fig2}.
The root-mean-square of $E-E_{\text{CK}}$ is 0.027~cm$^{-1}$. No significant systematic offset was found in comparison of level energies and transition wavelengths between values of the present work and of \cite{castelli2010new}, since $E-E_{\text{CK}} = 0$ is well within the spread of $E-E_{\text{CK}}$. Therefore, level energies and Ritz wavelengths newly determined in the present work are indeed at least twice and up to an order-of-magnitude more accurate compared to previously published values measured using stellar spectra of HR~6000. 

\subsection{Outlook}
Further improvements to wavelength accuracy and more Fe~II energy level identifications should be possible with more laboratory measurements. For example, the instrumental intensity response could be maximised and the bandwidth could be reduced using appropriate detectors and filters to target the 18,000 -- 21,000~cm$^{-1}$ region, increasing the observed S/Ns of transitions of the 4f levels. In Figure~\ref{fig1}, the noise in the Fe-Ne PDL FT spectrum FE1118.6 is visibly higher than that of the stellar spectrum, which explains the weaker lines used by \cite{castelli2010new} to identify Fe~II levels that were unobserved in the Fe-Ne PDL FT spectra. These levels have lines with relatively lower calculated transition probabilities \citep{kurucz2013fe} and an insufficient number of their lines were observed in the Fe-Ne PDL FT spectra for their definitive identifications. 

The greater S/Ns of transitions of high-lying levels in stellar absorption spectra compared to those in the laboratory emission spectra are explained by the relative population of the upper and lower levels, where lower levels are generally much better populated and so absorption spectroscopy could be advantageous. However, stellar spectra are more prone to blending due to many more elemental species in their plasmas. Hence, strategies for future analyses of the spectrum of Fe~II are expected to continue to involve corroboration between results from analysing laboratory and stellar spectra, as advantages of the two methods supplement each other.

\section{Summary}
FT spectroscopy of a Fe-Ne PDL in the visible spectral region has enabled 24 high-lying Fe~II energy levels to be determined in the laboratory for the first time, 20 of which belong to the 3d$^6$4f configuration and 4 of which belong to the 3d$^6$5d configuration. In total, 125 spectral lines corresponding to transitions of the 24 levels were observed. These levels were previously identified using only high-resolution stellar absorption spectra of HR~6000 \citep{castelli2010new}. Level energies and wavelength accuracies have been improved by at least a factor of two and up to one order-of-magnitude. Strategies for further investigation of the high-lying 4f energy levels of Fe~II were recommended to involve similar high-resolution FT spectroscopy of different detector-filter combinations and high-resolution stellar spectroscopy.

\section*{Acknowledgements}
The authors would like to thank the STFC and the Royal Society of the UK, and the bequest of Prof. Edward Steers for their support. H. K. thanks Dr Christian Clear for his support.

\section*{Data Availability}
The data underlying this paper will be shared on reasonable request
to the corresponding author.



\bibliographystyle{mnras}
\bibliography{example} 







\section*{SUPPORTING INFORMATION}
Full machine-readable form of Table~\ref{tab2} is available at ? online.

\bsp	
\label{lastpage}
\begin{landscape}
 \begin{table}
  \centering
  \caption{Classified transitions from the newly experimentally confirmed 4f and 5d energy levels of Fe~II to lower levels in the Fe-Ne PDL FT spectra (extract).}
  \label{tab3}
  \setlength{\tabcolsep}{2.3pt}
  \begin{tabular}{ccccccccccccccccc}
    \hline
     Upper Level$^a$ & $J_u$$^a$ & $E_u$$^a$ & Lower Level$^b$ & $J_l$$^b$ & $E_l$$^b$ & Sp.$^c$ & S/N & FWHM$^d$ & Int.$^d$ & $gA$$^a$ & $\log(gf)$$^a$ & $\sigma_{\text{cal}}$$^{d,e}$ & $\sigma_{\text{Ritz}}$$^e$ & $\sigma_{\text{cal}}$ - $\sigma_{\text{Ritz}}$ & $\lambda_{\text{Ritz}}^{\text{air}}$$^{e,f}$ & Comments$^g$\\
     & & (cm$^{-1}$) & & & (cm$^{-1}$) & & & (cm$^{-1}$) & (arb.) & ($10^6$ s$^{-1}$) & & (cm$^{-1}$) & (cm$^{-1}$) & (cm$^{-1}$) & (\AA{}) &\\
    \hline
    3d$^6$($^3$P2)4f 2[3]$^\circ$ & 7/2 & 122351.535 & 3d$^6$($^3$P2)4d $^2$F & 5/2 & 103597.409 & 2 & 11 & 0.148 & 160 & 773 & 0.518 & 18754.069(*) & 18754.126(25) & -0.057 & 5330.6772(71) & B Ne I \\
    3d$^6$($^3$P2)4f 2[3]$^\circ$ & 7/2 & 122351.535 & 3d$^6$($^3$P2)4d $^2$F & 7/2 & 103191.968 & 1 &  5 & 0.315 & 120 & 153 &-0.204 & 19159.543(*) & 19159.567(25) & -0.024 & 5217.8719(68) & B Fe I \\
    3d$^6$($^3$P2)4f 2[3]$^\circ$ & 7/2 & 122351.535 & 3d$^6$($^3$P2)4d $^4$F & 5/2 & 104569.226 & 2 &  4 & 0.203 &  35 &  94 &-0.349 & 17782.313(28)& 17782.309(24) &  0.004 & 5622.0060(76) & \\
    3d$^6$($^3$P2)4f 2[3]$^\circ$ & 7/2 & 122351.535 & 3d$^6$($^3$F2)4d $^2$D & 5/2 & 107407.818 & 2 &  3 & 0.243 &  35 &  68 &-0.343 & 14943.704(48)& 14943.717(25) & -0.013 & 6689.9285(112)& \\
    \hline
    3d$^6$($^3$P2)4f 2[5]$^\circ$ & 11/2 &122351.849 &  3d$^6$($^3$P2)4d $^4$F & 9/2 & 103165.320 & 1 & 26 & 0.249 & 590 &1525 & 0.793 & 19186.529(5) & 19186.529(5)  &  0.000 & 5210.5393(14) & \\
    3d$^6$($^3$P2)4f 2[5]$^\circ$ & 11/2 &122351.849 &  3d$^6$($^3$H)4d  $^4$F & 9/2 & 104916.552 & 2 &  4 & 0.075 &  16 &  94 &-0.333 & 17435.335(*) & 17435.297(9)  &  0.038 & 5733.9011(29) & W poor fit\\
    \hline
    $\vdots$ &$\vdots$ &$\vdots$ &$\vdots$ &$\vdots$ &$\vdots$ &$\vdots$ &$\vdots$ &$\vdots$ &$\vdots$ &$\vdots$ &$\vdots$ &$\vdots$ \\
    \hline
    3d$^6$($^3$H)5d $^4$I$^\circ$ & 15/2 & 124357.282 & 3d$^6$($^3$H)5p $^4$I$^\circ$ & 13/2 & 109619.704 &2&31&0.191& 360 & 492 & 0.531 & 14737.579(3) & 14737.578(3)  &  0.001 & 6783.5032(15) & \\
    3d$^6$($^3$H)5d $^4$I$^\circ$ & 15/2 & 124357.282 & 3d$^6$($^3$H)5p $^4$I$^\circ$ & 15/2 & 109591.643 &2&10&0.222& 130 & 350 & 0.381 & 14765.640(*) & 14765.639(7)  &  0.001 & 6770.6116(30) & B Fe III \\
    3d$^6$($^3$H)5d $^4$I$^\circ$ & 15/2 & 124357.282 & 3d$^5$($^4$G)4s4p($^1$P$^\circ$) $^4$H$^\circ$ & 13/2&108729.162&2&13&0.189&140&313&0.284&15628.115(7)&15628.120(6)&-0.005&6396.9540(23) & \\
    3d$^6$($^3$H)5d $^4$I$^\circ$ & 15/2 & 124357.282 & 3d$^5$($^2$G)4s4p($^3$P$^\circ$) $^4$H$^\circ$ & 13/2&110348.401&2& 5&0.236&84 &168&0.108&14008.909(23)&14008.881(6)&0.028&7136.3616(33) & \\
    3d$^6$($^3$H)5d $^4$I$^\circ$ & 15/2 & 124357.282 & 3d$^5$($^2$I)4s4p($^3$P$^\circ$) $^2$I$^\circ$ & 13/2&109149.611&2&11&0.196&130&132&-0.069&15207.696(*)&15207.671(7)&0.025&6573.8130(28) & B U \\
    3d$^6$($^3$H)5d $^4$I$^\circ$ & 15/2 & 124357.282 & 3d$^6$($^3$H)5p $^4$H$^\circ$ & 13/2&111651.055&2& 4&0.113&  76&129&0.077&12706.148(*)&12706.227(7)&-0.079&7867.9923(40)&W + B U\\
    \hline
    $\vdots$ &$\vdots$ &$\vdots$ &$\vdots$ &$\vdots$ &$\vdots$ &$\vdots$ &$\vdots$ &$\vdots$ &$\vdots$ &$\vdots$ &$\vdots$ &$\vdots$ \\
    \hline
    3d$^6$($^3$G)4f 4[4]$^\circ$ & 9/2 & 127869.923 & 3d$^6$($^3$G)4d  $^4$G & 7/2 & 108537.623 & 1 & 11 & 0.247 & 190 & 487 & 0.291 & 19332.305(11)& 19332.300(8)  & 0.005  & 5171.2499(22) & B Fe I resolved\\
    3d$^6$($^3$G)4f 4[4]$^\circ$ & 9/2 & 127869.923 & 3d$^6$($^3$G)4d  $^4$H & 9/2 & 108577.597 & 1 &  4 & 0.284 &  88 & 323 & 0.114 & 19292.313(65)& 19292.326(9)  &-0.013  & 5181.9650(24) & \\
    3d$^6$($^3$G)4f 4[4]$^\circ$ & 9/2 & 127869.923 & 3d$^6$($^3$G)4d  $^2$H & 9/2 & 110008.310 & 2 &  6 & 0.194 &  55 & 227 & 0.029 & 17861.601(17)& 17861.613(9)  &-0.012  & 5597.0446(27) & \\
    3d$^6$($^3$G)4f 4[4]$^\circ$ & 9/2 & 127869.923 & 3d$^6$($^3$G)4d  $^2$F & 7/2 & 110570.312 & 2 &  6 & 0.140 &  61 & 175 &-0.058 & 17299.612(11)& 17299.611(8)  & 0.001  & 5778.8742(27) & \\
    3d$^6$($^3$G)4f 4[4]$^\circ$ & 9/2 & 127869.923 & 3d$^6$($^3$G)4d  $^2$G & 7/2 & 109901.523 & 2 &  3 & 0.044 &  10 & 107 &-0.303 & 17968.362(*) & 17968.400(9)  &-0.038  & 5563.7807(28) & W poor fit \\
    \hline
    3d$^6$($^3$G)4f 4[6]$^\circ$ & 11/2 & 127880.453 & 3d$^6$($^3$G)4d  $^4$H & 9/2 & 108577.597 & 1 & 15 & 0.221 & 300 & 752 & 0.481 & 19302.853(7) & 19302.856(7)  &-0.003  & 5179.1381(20) & \\
    3d$^6$($^3$G)4f 4[6]$^\circ$ & 11/2 & 127880.453 & 3d$^6$($^3$G)4d  $^4$H & 11/2& 108387.934 & 1 & 10 & 0.283 & 240 & 289 & 0.057 & 19492.499(*) & 19492.519(9)  &-0.020  & 5128.7443(24) & B Fe II \\
    3d$^6$($^3$G)4f 4[6]$^\circ$ & 11/2 & 127880.453 & 3d$^6$($^3$G)4d  $^4$G & 9/2 & 108391.507 & 1 &  3 & 0.345 & 120 & 265 & 0.020 & 19489.079(*) & 19488.946(9)  & 0.133  & 5129.6845(24) & W + B Fe I \\
    3d$^6$($^3$G)4f 4[6]$^\circ$ & 11/2 & 127880.453 & 3d$^6$($^3$G)4d  $^2$G & 9/2 & 109625.216 & 1 &  2 & 0.088 &  20 & 220 &-0.005 & 18255.193(*) & 18255.237(9)  &-0.044  & 5476.3584(28) & B Fe I \\
    3d$^6$($^3$G)4f 4[6]$^\circ$ & 11/2 & 127880.453 & 3d$^6$($^3$G)4d  $^2$H & 9/2 & 110008.310 & 2 &  6 & 0.202 &  55 &  95 &-0.352 & 17872.161(18)& 17872.143(9)  & 0.018  & 5593.7468(28) & \\
    3d$^6$($^3$G)4f 4[6]$^\circ$ & 11/2 & 127880.453 & 3d$^6$($^3$G)4d  $^4$I & 11/2& 108775.110 & 1 &  2 & 0.051 &   9 &  86 &-0.450 & 19105.393(*) & 19105.343(9)  & 0.050  & 5232.6812(25) & B U \\
    3d$^6$($^3$G)4f 4[6]$^\circ$ & 11/2 & 127880.453 & 3d$^6$($^3$G)4d  $^2$I & 11/2& 109389.879 & 2 &  2 &     - &   - &  73 &-0.493 & 18490.570(*) & 18490.574(10) &-0.004  & 5406.6578(29) & W not fitted \\
    \hline
    \multicolumn{17}{l}{The full version of table is available online.} \\
    \multicolumn{17}{l}{$^a$ Upper level labels and transition probabilities from \citet{kurucz2013fe}, upper level energies $E_u$ are optimised values of the present work.} \\
    \multicolumn{17}{l}{$^b$ Lower level labels from \citet{kramida2023nist} and energies from \citet{nave2012spectrum}, lower level energies $E_l$ were fixed in the level optimisation.} \\
    \multicolumn{17}{l}{$^c$ 1 for lines from spectrum FE1118.6 and 2 for lines from spectrum FE1120.2.} \\
    \multicolumn{17}{l}{$^d$ FWHM and relative intensity are `-' for spectral lines that are not fitted due to insufficient S/N for reliable Voigt profile fitting, observed wavenumbers are determined using peak values.} \\
    \multicolumn{17}{l}{$^e$ Observed and calibrated wavenumber $\sigma_{\text{cal}}$, Ritz wavenumber $\sigma_{\text{Ritz}}$, and Ritz air wavelength $\lambda_{\text{Ritz}}^{\text{air}}$. Standard uncertainties are in parentheses in units of the final decimal place, `*' indicates the observed}\\
    \multicolumn{17}{l}{\:\:\:\,wavenumber is neglected in level optimisation due to line blending and/or low S/N.}\\
    \multicolumn{17}{l}{$^f$ Converted using the three term dispersion formula from \citet{peck1972dispersion}.} \\
    \multicolumn{17}{l}{$^g$ `B' is for blend and `W' is for weak, unknown species are indicated by `U'.}
  \end{tabular}
 \end{table}
\end{landscape}
\end{document}